\documentclass[3p,times,procedia]{elsarticle}
\usepackage{nupha_ecrc}
\usepackage{lineno}
\usepackage{amssymb}
\usepackage{amsmath}
\usepackage{wrapfig}
\usepackage{graphicx}
\usepackage{verbatim}

\volume{00}
\firstpage{1}
\journalname{Nuclear Physics A}
\runauth{}
\jid{nupha}
\jnltitlelogo{Nuclear Physics A}

\begin{document}

\begin{frontmatter}
\dochead{}
\title{Beam-Energy and Centrality Dependence of Directed Flow of  Identified Particles }
\author{Prashanth Shanmuganathan (for the STAR Collaboration)\footnote{A list of members of the STAR Collaboration and acknowledgements can be found at the end of this issue.}}
\address{Department of Physics, Kent State University, Kent, OH 44242}

\begin{abstract}
These proceedings present directed flow ($v_1$) measurements in Au+Au collisions from STAR's Beam Energy Scan (BES) program at the Relativistic Heavy-Ion Collider, for $p$, $\bar{p}$, $\Lambda$, $\bar\Lambda$, $K^\pm$, $K^0_S$ and $\pi^\pm$. At intermediate centrality, protons show a minimum in directed flow slope, $dv_1/dy\,|_{y\leq0.8}$, as a function of beam energy. Proton $dv_1/dy$ changes sign near 10 GeV, and the directed flow for $\Lambda$ is consistent with the proton result. The directed flow slope for net protons shows a clear minimum at 14.5 GeV and becomes positive at beam energies below 10 GeV and above 30 GeV. New results for net-kaon directed flow slope resemble net protons from high energy down to 14.5 GeV, but remain negative at lower energies. The slope $dv_1/dy$ shows a strong centrality dependence, especially for $p$ and $\Lambda$ at the lower beam energies. Available model calculations are in poor agreement. 

\end{abstract}


\end{frontmatter}


\section{Introduction}
The Relativistic Heavy-Ion Collider has explored Au+Au collisions over the range $\sqrt{s_{NN}} = 7.7$ to 200 GeV, normally referred to as the Beam Energy Scan (BES). At the full energy of RHIC, there is evidence of a smooth crossover from hadronic matter to quark matter, whereas additional features of the QCD phase diagram, like a possible first-order phase transition and a critical point, may come into play as the lower BES range is explored \cite{whitepapers,BES-II}.  

The observed distribution of azimuths ($\phi$) relative to the reaction plane ($\Psi_{\rm RP}$) is described using a Fourier expansion. The first Fourier coefficient is called directed flow, $v_1 = \langle\cos(\phi - \Psi_{\rm RP})\rangle$ \cite{methods}. The slope $dv_1/dy$ near midrapidity is a useful observable to characterize the overall strength of directed flow when studying its dependence on beam energy, particle species, centrality, etc. Previous proton $dv_1/dy$ measurements for 10-40\% centrality from STAR show a sign change and a minimum in the beam energy dependence, whereas antiprotons and charged pions show monotonic behavior throughout $\sqrt{s_{NN}} = 7.7$ to 200 GeV and remain negative for all available energies \cite{STAR-BESv1}. In earlier hydrodynamic calculations, a minimum in directed flow was predicted when the equation of state incorporated a first-order phase transition, while a purely hadronic equation of state predicted a monotonic beam energy dependence \cite{Rischke,Stoecker}. However, more recent model calculations with and without phase transitions show only monotonic behavior at BES energies and strongly disagree with the published STAR proton data \cite{hybrid,phsd,phsd2}. The new preliminary $v_1$ measurements presented here for $\Lambda$, $\bar\Lambda$, $K^\pm$ and $K^0_S$ at BES energies offer many additional phenomenological details, and are expected to constrain the role of the QCD equation of state in future model descriptions. 

\section{Analysis Details}
The STAR detector has full azimuthal coverage over pseudorapidity $|\eta|$$<$1, with good particle ID for charged  pions, kaons and protons via $dE/dx$ \cite{startpc} and time of flight \cite{TOF}. The neutrals $\Lambda$ and $K^0_s$ are reconstructed using the invariant mass method, identifying decay daughters either using information from the STAR Time Projection Chamber alone with tight topological cuts, or using broader topological cuts in cases where daughter tracks have TOF hits \cite{STAR-BESv2}. The mixed-event method is employed to subtract the combinatoric background from the yield \cite{STAR-BESv2}. 
Pions and kaons were required to have transverse momentum $p_T$$>$0.2 GeV$/c$ and momentum $p$$<$1.6 GeV$/c$. The studied protons and antiprotons have $p_T$ between 0.4 and 2.0 GeV$/c$ and $\Lambda$ and $K^0_s$ have $p_T$ between 0.2 and 5.0 GeV$/c$. The first-order event plane azimuth is estimated using the STAR Beam-Beam Counters (BBC) \cite{BBC}, which cover 3.3$<$$|\eta|$$<$5.0. This range leaves a substantial gap in $\eta$ between the analyzed particles and those used for event plane determination, which helps to suppress non-flow effects \cite{methods,CuCuPaper,STAR-BESv1}. 

\section{Main Findings}
\begin{figure}[h]
\center
\includegraphics[width=0.99\textwidth]{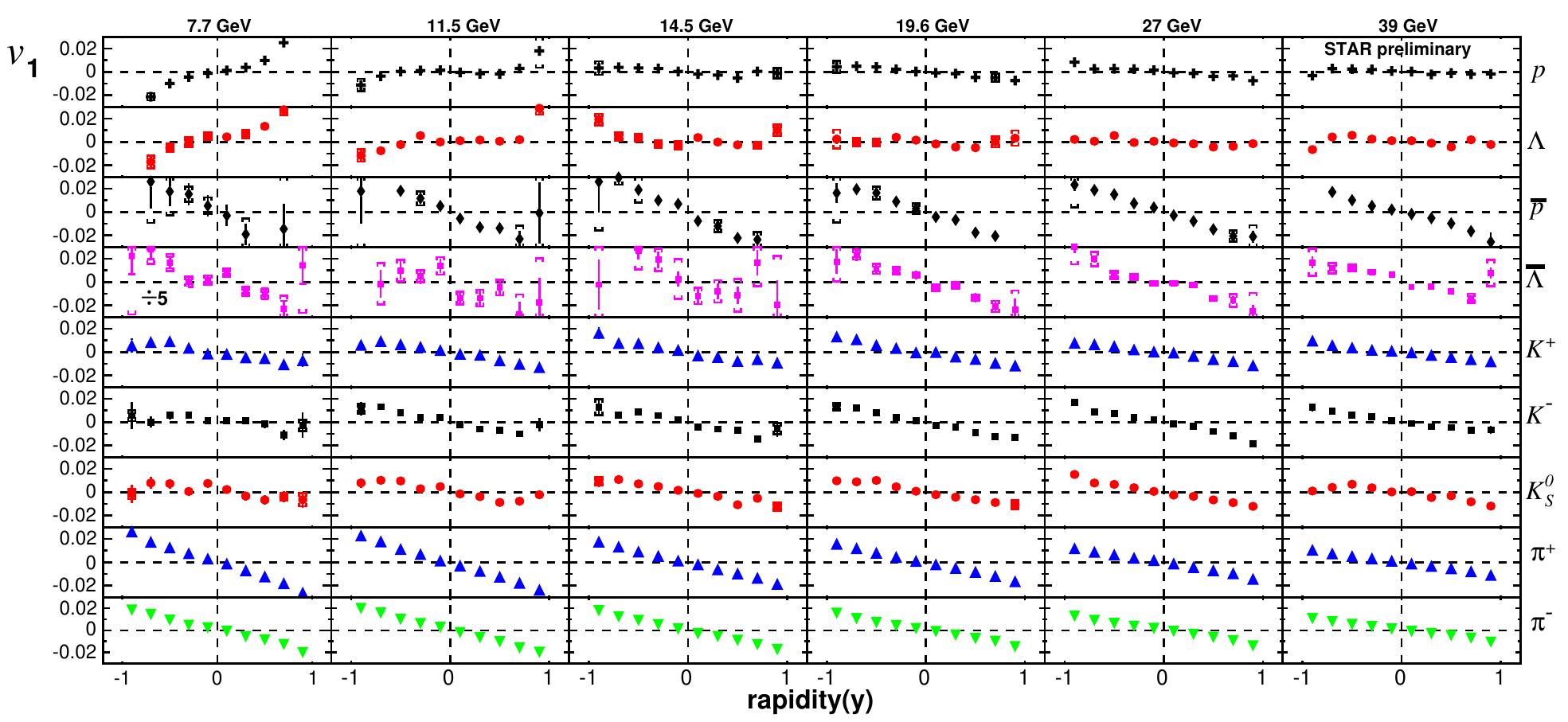}
\caption{Au + Au directed flow ($v_1$) plotted as a function of rapidity for protons, $\Lambda$, antiprotons, $\bar\Lambda$, $K^\pm$, $K^0_s$ and $\pi^\pm$, at 10-40\% centrality. Error bars are statistical, while caps indicate systematic errors. Note that the $v_1$ magnitude is exceptionally large for $\bar\Lambda$ at 7.7 GeV, and therefore for that panel only, $v_1$ and its errors are divided by 5 to fit on the common vertical scale. }
\label{v1vsy}
\end{figure}

\begin{wrapfigure}{R}{0.5\textwidth}
\includegraphics[height=0.53\textwidth]{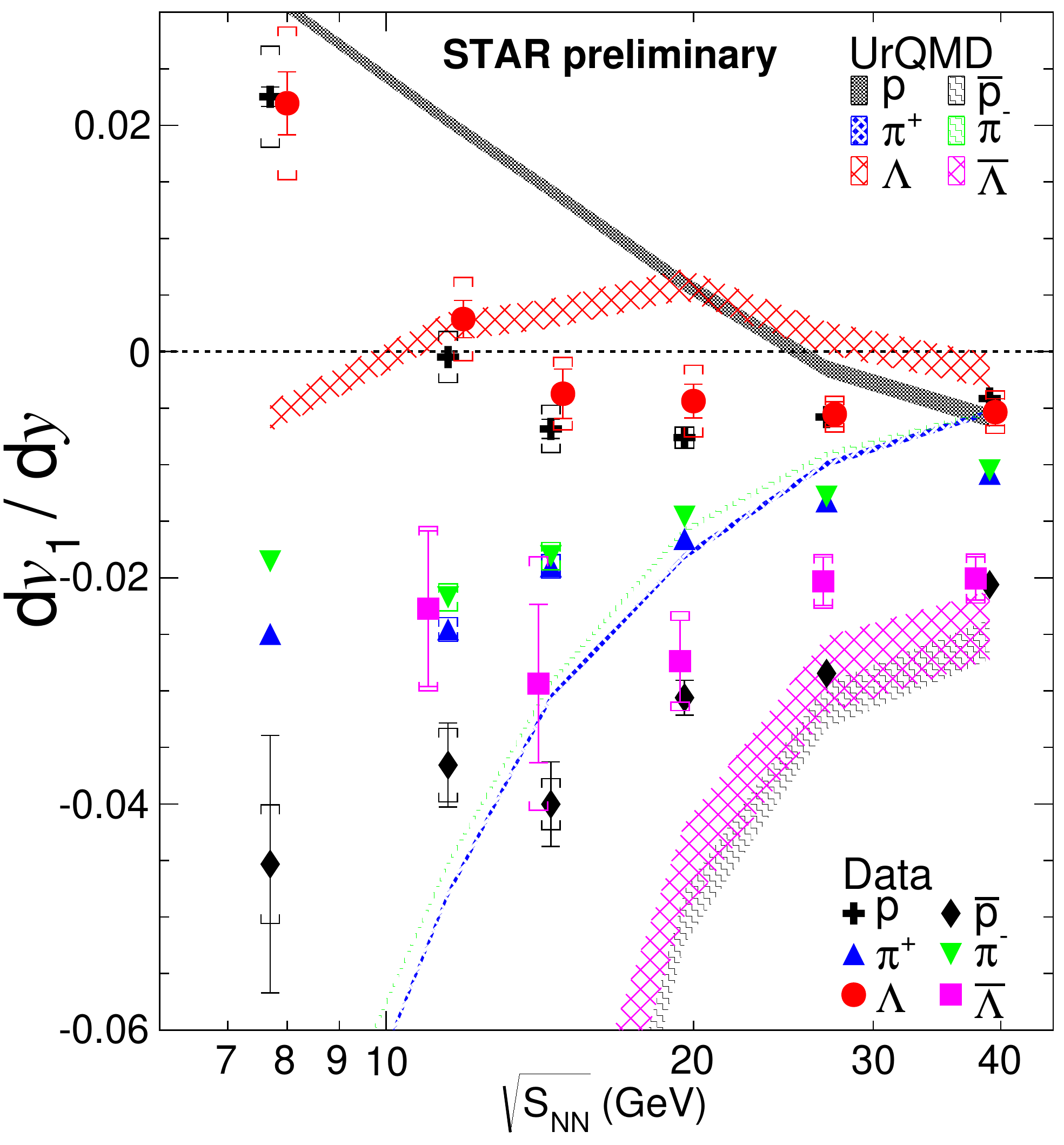}
\caption{Slope of directed flow ($dv_1/dy$) near midrapidity as function of beam energy for protons, antiprotons, $\pi^\pm$, $\Lambda$ and $\bar\Lambda$ in 10-40\% centrality Au+Au collisions. Shaded bands show UrQMD calculations. The $dv_1/dy$ for $\bar\Lambda$ at 7.7 GeV is off-scale, well below the lower end of the vertical axis. Error bars are statistical, while caps indicate systematic errors. }
\label{dv1dy1}
\end{wrapfigure}

Figure \ref{v1vsy} presents $v_1(y)$ measurements for nine particle species produced in Au+Au collisions at 10-40\% centrality.  Some of these results were published in Ref.~\cite{STAR-BESv1}, namely protons, antiprotons and charged pions at $\sqrt{s_{NN}} = 7.7$, 11.5, 19.6, 27 and 39 GeV. Since the new beam energy of $\sqrt{s_{NN}} = 14.5$ GeV is now available, the full set of panels for protons, antiprotons and charged pions is shown in Fig.~\ref{v1vsy} for completeness.  

Figure \ref{dv1dy1} shows the slope $dv_1/dy$ near midrapidity as a function of beam energy for $p$, $\bar{p}$, $\pi^\pm$,  $\Lambda$, and $\bar\Lambda$ in 10-40\% centrality Au+Au collisions. 
Previous measurements of this type were based on the linear term $dv_1/dy$ in a cubic fit to $v_1(y)$ \cite{STAR-BESv1}. A cubic fit somewhat reduces sensitivity to the rapidity range over which the fit is performed, but a cubic fit becomes unstable when statistics are poor, as is now the case for $\bar{\Lambda}$, and to a lesser extent, $\Lambda$. Therefore, in the present analysis, a linear fit over $|y| \leq 0.8$ has been used for all particle species at all beam energies. 

A noteworthy observation is that $dv_1/dy$ for $p$ and $\Lambda$ agree within the uncertainties, and both species changes sign around the same beam energy. Protons show a minimum in $dv_1/dy$ near 15 GeV, whereas $\Lambda$ errors are not small enough to draw a conclusion about whether the minimum in proton slope also occurs for $\Lambda$. The slopes for $\bar{p}$, $\bar\Lambda$ and $\pi^\pm$ are negative for all beam energies, and $\bar{p}$ and $\bar\Lambda$ are roughly consistent within errors. UrQMD \cite{urqmd} calculations are drastically different from the experiment for $p$ and $\Lambda$, and unlike the data, show a simple monotonic behavior. However, UrQMD shows qualitatively similar trends as data for $\bar{p}$, $\bar\Lambda$ and $\pi^\pm$. 

\begin{figure*}[h]
    \centering
    \begin{minipage}[b]{.6\linewidth}
        \centering
        \includegraphics[width=1.02\textwidth]{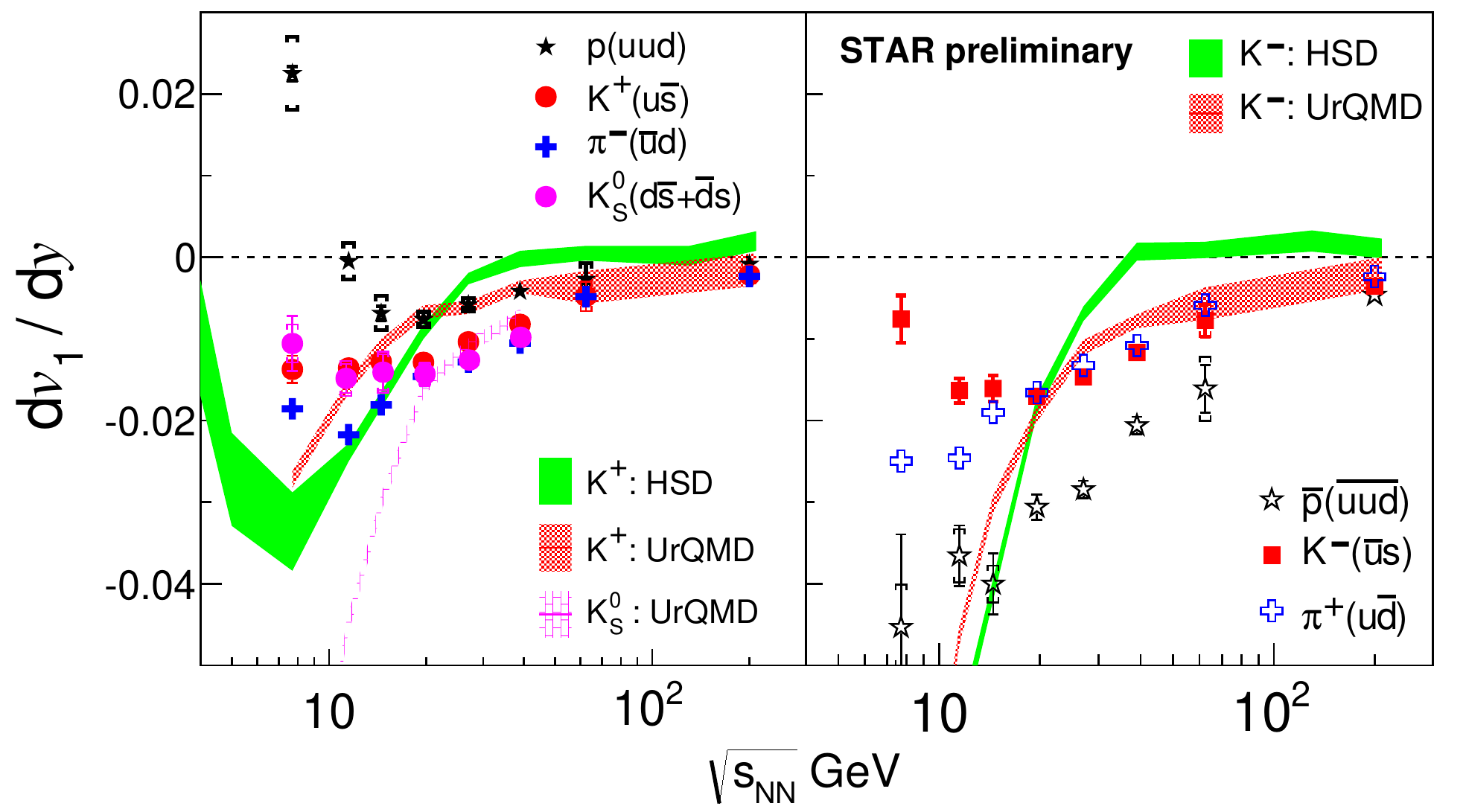}
    \end{minipage}%
    \hfill%
    \begin{minipage}[b]{.4\linewidth}
    \centering
        \centering
        \includegraphics[width=1.0\textwidth]{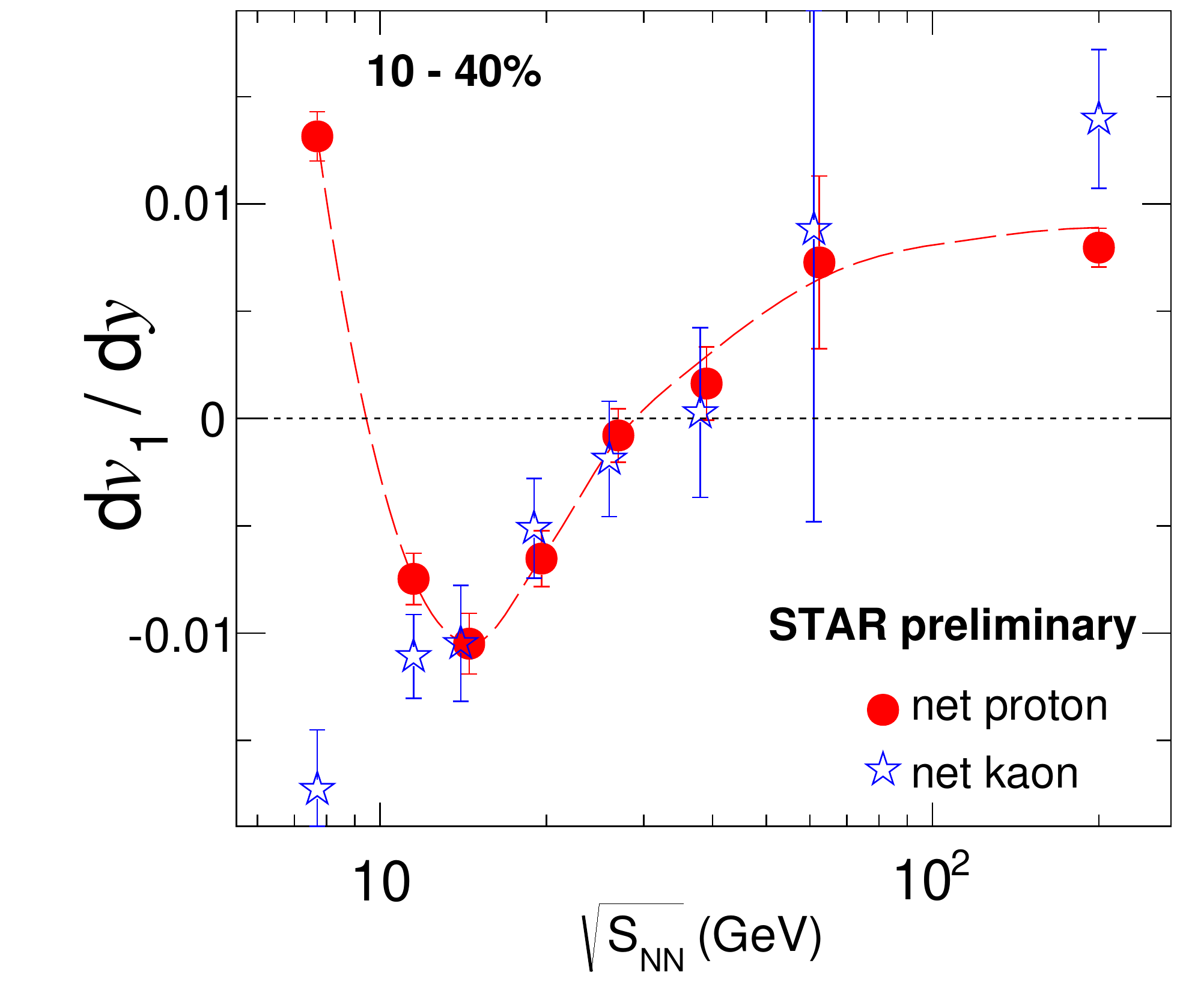}
    \end{minipage}\\[-8pt]
    \begin{minipage}[t]{.59\linewidth}
        \caption{Au+Au $dv_1/dy$ near mid-rapidity at 10-40\% centrality for $p$, $K^+$, $\pi^-$ and $K^0_s$ in the panel on the left, and for $\bar p$, $K^-$ and $\pi^+$ on the right. Error bars are statistical, while caps indicate systematic errors. UrQMD calculations are shown for all kaon charges, and HSD is shown for $K^\pm$.}
                \label{dv1dy2}
    \end{minipage}%
    \hfill%
    \begin{minipage}[t]{.37\linewidth}
        \caption{Au+Au $dv_1/dy$ near mid-rapidity at 10-40\% centrality for net protons and net kaons. The dashed curve is a smooth fit to the net-$p$ data points to guide the eye.}
                \label{netp}
    \end{minipage}%
\end{figure*}

Figure \ref{dv1dy2} focuses on $dv_1/dy$ for charged and neutral kaons, with $p$, $\bar{p}$ and $\pi^\pm$ shown again as a reference.  Particles ($p$, $K^+$, $K^0_s$ and $\pi^-$) are grouped together in the panel on the left, while the corresponding antiparticles are compared in the panel on the right. In other words, we expect to find more quarks from stopped initial-state nucleons in the left panel than in the right panel. The $dv_1/dy$ for $K^+$, $K^-$ and $K^0_s$ all are negative for all BES energies, and $dv_1/dy$ for $K^0_s$ is consistent with the mean of $K^+$ and $K^-$ within errors.  Kaon calculations from Hadron String Dynamics (HSD) \cite{phsd,phsd2} and UrQMD \cite{urqmd} are also compared (shaded bands) and show qualitatively similar trends as data at higher energies, but deviate more strongly at lower energies.
  
In Fig.~\ref{netp}, a comparison is shown between net-proton $dv_1/dy$ and net-kaon $dv_1/dy$.
Net-proton $v_1$ is defined according to $\lbrack v_1(y)\rbrack_p = r(y)\lbrack v_1(y)\rbrack_{\bar{p}} + (1- r(y))\, \lbrack v_1(y)\rbrack_{{\rm net\mbox{-}}p}$, where $r(y)$ is the observed rapidity dependence of the ratio of antiprotons to protons. Net-kaon $v_1$ is defined using the same formula, with $p$ and $\bar p$ replaced by $K^+$ and $K^-$, respectively. This net-$p$ definition attempts to distinguish between protons from baryon-antibaryon pairs produced near midrapidity, as opposed to protons arising from initial-state baryon number transported to the midrapidity region by stopping. This distinction in turn is based on the hypothesis that antiprotons and protons from produced quarks have roughly the same directed flow. Final-state interaction effects, such as annihilation and hadronic potentials, complicate the
simplified picture above. It is evident that net-$p$ and net-$K$ $dv_1/dy$ near midrapidity are consistent with each other from 200 GeV down to 14.5 GeV, begin to diverge at 11.5 GeV, and are dramatically different at 7.7 GeV.     

\begin{figure}
\center
\includegraphics[width=1.0\textwidth]{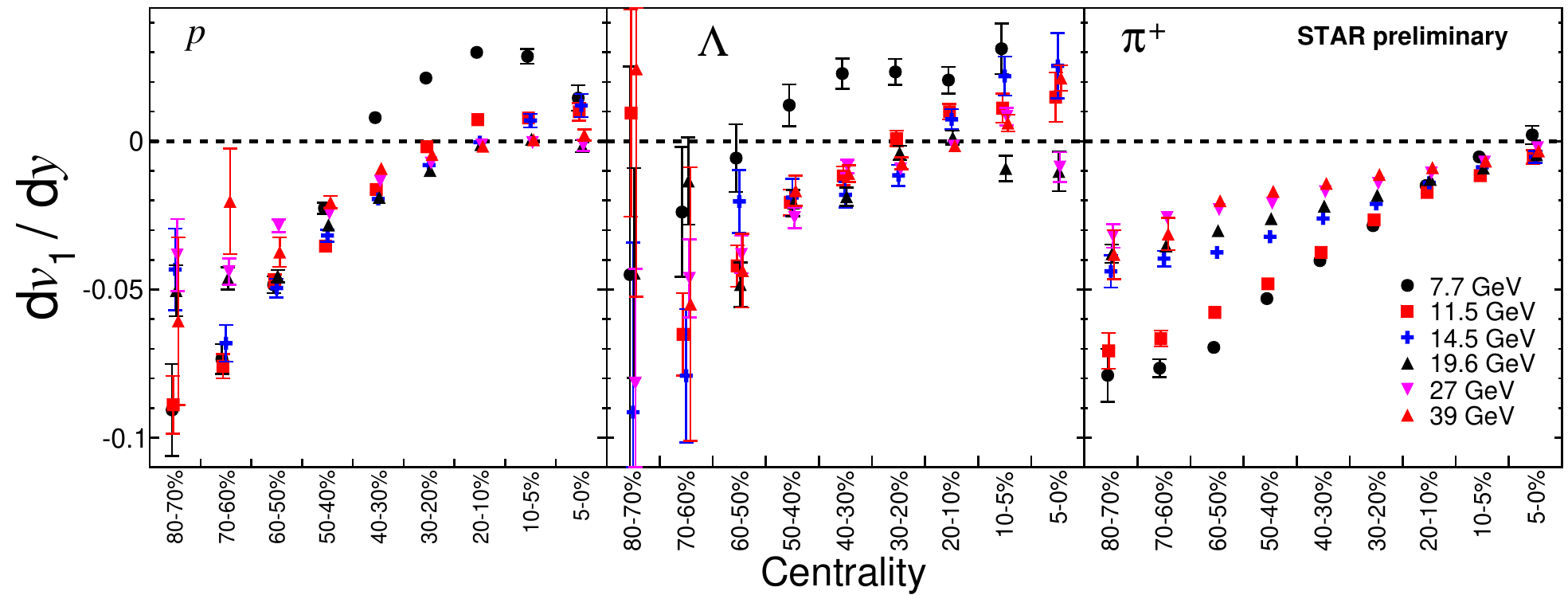}
\caption{Slope of directed flow ($dv_1/dy$) near midrapidity as a function of centrality for $p$, $\Lambda$ and $\pi^+$ in Au+Au collisions. Error bars are statistical only; systematic errors are generally comparable to statistical errors, but are omitted from these panels to reduce clutter.}
\label{v1vscent}
\end{figure}

Figure \ref{v1vscent} shows the centrality dependence of directed flow slope near midrapidity for $p$, $\Lambda$ and $\pi^+$.  The corresponding result for $\pi^-$ is very similar to $\pi^+$ and is not shown. Kaons, also not shown, are qualitatively similar to pions but with larger uncertainties.  Statistics for $\bar{p}$ and $\bar{\Lambda}$ are insufficient for a centrality dependent study. The centrality dependence for $p$, $\Lambda$ and $\pi^\pm$ is very strong, with $p$ and $\Lambda$ showing a sign change as centrality is varied while holding beam energy fixed at many of the BES points. 

Overall, it is evident that there is a rich and detailed structure present in the beam energy and centrality dependence of directed flow for the nine particle species included in the present study. These measurements already pose a considerable challenge to existing models. Much more comprehensive measurements in the region of high baryon density will be possible with the greatly increased statistics of Phase-II of the RHIC Beam Energy Scan \cite{BES-II}.

\end{document}